\documentclass[twocolumn,superscriptaddress,preprintnumbers]{revtex4-1}

\usepackage{amssymb,amsmath,latexsym,bm,amsfonts}

\usepackage{longtable}
\usepackage{color,xcolor}
\usepackage{indentfirst}
\usepackage[margin=1in]{geometry}
\usepackage{mathtools,tensor,graphicx,}

\usepackage{stmaryrd}
\usepackage[colorlinks=true,filecolor=red,citecolor=blue ]{hyperref}
\usepackage{cleveref} 
\usepackage{eufrak}
 \usepackage{mathrsfs}

\usepackage{tikz,colortbl}
\usetikzlibrary{calc}
\usepackage{zref-savepos}
\usepackage{array}

\numberwithin{equation}{section}

\begin{document}
\preprint{YITP-22-28, IPMU22-0013}
\title{Self-tuning of the cosmological constant in brane-worlds with $P(X,\phi)$}
\author{Osmin Lacombe}\email{lacombe.osmin@yukawa.kyoto-u.ac.jp}
\affiliation{Center for Gravitational Physics and Quantum Information, Yukawa Institute for Theoretical Physics, Kyoto University, Kyoto, 606-8502, Japan}
\author{Shinji Mukohyama}\email{shinji.mukohyama@yukawa.kyoto-u.ac.jp}
\affiliation{Center for Gravitational Physics and Quantum Information, Yukawa Institute for Theoretical Physics, Kyoto University, Kyoto, 606-8502, Japan}
\affiliation{Kavli Institute for the Physics and Mathematics of the Universe (WPI), The University of Tokyo Institutes for Advanced Study, The University of Tokyo, Kashiwa, Chiba, 277-8583, Japan}
\date{\today}
\begin{abstract}
We revisit the idea of self-tuning the observed cosmological constant to a vanishing value and promote it to a selection criterion of brane-world models, in which our Universe is described by a $3$-brane embedded in a $5d$ bulk. As a concrete setup, we consider a bulk scalar field $\phi$ described by a general Lagrangian $P(X,\phi)$ with $X=-(\partial\phi)^2/2$. By requiring that the model enforces the $4d$ curvature of the maximally symmetric $3$-brane world-volume to vanish independently of the $4d$ effective vacuum energy, only two possibilities remain: one with a canonical bulk kinetic term and the other with an unconventional bulk kinetic term similar to a Cuscuton field. Further demanding the absence of bulk singularity, the latter is selected as a unique possibility within the class of models. {At the background level, the} solution can accommodate any warp factor profile free from bulk singularity and with a finite effective $4d$ Planck mass. In a cosmological context, our solution would describe our (almost) flat Universe at late times, with a bulk warp factor profile expected to be determined by the evolution of the Universe before dilution of the matter fields by cosmic expansion.  {Eventually, a simple analysis is performed in the bulk showing no obvious instability around the background solution. A full stability analysis taking into account brane bending modes is nevertheless necessary and left for future work.}
\end{abstract}

\maketitle

\section{Introduction}

The cosmological constant problem is usually described as the discrepancy between the (almost) vanishing observed curvature of our Universe and the large contributions to the vacuum energy coming from the Standard Model sector. Indeed, from the gravity point of view, the effective vacuum energy of our Universe can be described by a non-vanishing cosmological constant related to the curvature of space-time. The observed curvature of our Universe hence describes an almost vanishing effective vacuum energy, which would imply an extreme fine-tuning of the bare cosmological constant value to cancel the large contributions from the Standard Model quantum corrections. 

The introduction of extra dimensions is an elegant way to ameliorate the cosmological constant problem. Solutions where the effective vacuum energy only curves unobserved (extra) dimensions, leading to a seemingly vanishing cosmological constant, were found originally in \cite{Rubakov:1983bb,Rubakov:1983bz}. However, no satisfactory mechanism was found to select these solutions among other solutions where the observed dimensions are curved as well. More than a decade after these original considerations, the possibility for extra dimensions and brane-world scenarios to ameliorate the hierarchy problem \cite{Antoniadis:1998ig,Kaloper:1999sm,Randall:1999ee,Randall:1999vf} also lead to great progress toward a possible solution to the cosmological constant problem. Brane-world scenarios consider that Standard Model fields live on a thin brane, identified to our observed Universe, embedded in a higher-dimensional space-time. {Several authors \cite{Arkani-Hamed:2000hpr,Csaki:2000wz,Kachru:2000hf,Forste:2000ft} realized that extra dimensions are elegant ways to evade Weinberg's no-go theorem \cite{Weinberg:1988cp} on adjustment mechanisms of the cosmological constant (see also \cite{Niedermann:2017cel} and references therein for discussions on various further obstructions). In particular, these authors} developed models where the curvature of the brane vanishes independently of its effective vacuum energy, introducing the notion of self-tuning.  However, the solutions found in these models suffered from bulk singularities at finite distances (which cannot be shielded by a horizon \cite{Cline:2001yt}) and it was shown that resolving them through additional branes would amount to reintroducing fine-tuning \cite{Forste:2000ps}. 

After these initial works on self-tuning mechanisms, attempts to avoid bulk singularities were developed using peculiar bulk scalar fields, described in terms of unconventional kinetic terms \cite{Forste:2011hq} or yet equation of states \cite{Antoniadis:2010ik,Antoniadis:2021grg,Antoniadis:2021rxw}. Such fields can be described in terms of k-essence scalars \cite{Garriga:1999vw,Armendariz-Picon:1999hyi},  a general class of scalar fields with higher-order kinetic terms. Originally introduced for inflationary model building, k-essence scalars constitute nowadays standard building blocks of modified gravity theories. Ultraviolet completions of k-essence models, including supersymmetric extensions, were investigated in the past \cite{Khoury:2010gb,Farakos:2012je,Farakos:2012qu,Koehn:2012ar}. 

Self-tuning mechanisms without bulk singularities were also developed in the past decades in the holographic context, through asymmetric brane-world models \cite{Amariti:2019vfv,Charmousis:2017rof,Hamada:2020bbf}. For the sake of completeness, we also refer here to alternative ways to address the cosmological constant problem in four-dimensional cosmological models -- through sequestration \cite{Kaloper:2013zca,Kaloper:2014fca,Kaloper:2015jra,Kaloper:2016yfa} or dynamical cancellation of the vacuum energy in general scalar-tensor theories \cite{Mukohyama:2003nw,Charmousis:2011ea,Charmousis:2011bf,Appleby:2012rx,Appleby:2018yci,Copeland:2021czt,Khan:2022bxs}, {as well as six-dimensional supersymmetric constructions  \cite{Aghababaie:2003wz}. }

In this paper, inspired by the aforementioned works exploring ways to avoid brane-worlds bulk singularities with unconventional bulk fields, we study the case of generic k-essence bulk scalars. We start with a general description of these setups. This class of model allows reconstructing any (regular enough) warp factor profile, at the background level, in the spirit of the cosmological Hubble factor reconstructions realized in modified gravity effective theories \cite{Creminelli:2006xe,Pirtskhalava:2014esa,Alberte:2016izw}. Nevertheless, asking for self-tuning of the observed brane cosmological constant, namely enforcing a flat brane solution, greatly restrict the choice of brane coupling and k-essence defining functions. The main result of this paper follows from this consideration and resides in the discovery of a very simple self-tuning solution described by a Cuscuton \cite{Afshordi:2006ad,Afshordi:2007yx} bulk scalar. This solution can accommodate any warp factor profile, in particular regular ones with a finite effective $4d$ Planck mass. A simple analysis suggests that it does not suffer from obvious instability issues in the bulk.

The paper is organized as follows. In \cref{section:kessencebulk}, we describe our brane-world setup in presence of a generic k-essence bulk scalar and introduce our notations, before deriving the background equations and brane junction conditions in \cref{section:bckgrdandjunctions}. In \cref{section:selftuning}, we recall the essence of the cosmological constant self-tuning mechanism and analyze its requirements in presence of a k-essence bulk scalar. We find a new possible form for the bulk scalar kinetic term, namely a specific k-essence function $P$, implementing the self-tuning mechanism for an exponential brane scalar potential. We note that the brane scalar potential can always be chosen in this form, by redefinition of the scalar field, and thus inspect in more detail this general new solution corresponding to a Cuscuton bulk scalar. In \cref{section:stability}, we eventually comment on the causality and stability of our solution, before discussing in \cref{section:discussion} natural continuations to this work. 
The paper also includes two appendices, rather independent of the main results of the paper: \cref{appendix:A} shows how to relate brane-worlds with k-essence bulk scalars with other models studied in the literature, and \cref{appendix:B} describes the scalar field junction conditions  in the particular case where the k-essence describes a non-linear bulk fluid. 

\section{Brane-world model with a bulk k-essence scalar} \label{section:kessencebulk}

In this work, we consider a simple brane-world model where our Universe is described by the 1+3 dimensional world-volume of a 3-brane embedded in a $5d$ bulk spacetime. We will be interested in solutions with maximally-symmetric brane geometries, and choose our coordinate system so that the brane is orthogonal to the bulk fifth direction $y$. {We take a non-compact fifth dimension, as in the original setup of \cite{Randall:1999vf}, so that the $y$ coordinates runs {\it a priori} from $y=-\infty$ to $y=+\infty$. We emphasize that a non-compact fifth dimension, leading to an ungapped Kaluza-Klein spectrum, does not lead to the standard $4d$ description of gravity, and hence evade Weinberg's no-go theorem on the tuning of the cosmological constant  \cite{Weinberg:1988cp}. As noticed in  \cite{Randall:1999vf}, a non-compact extra dimension can nevertheless lead to localized gravity.}

{We take the position of the brane in the fifth direction as the origin of the $y$ coordinate,} namely we take it at $y=0$. In the following, we consider solutions with a mirror $\mathbb Z_2$ symmetry in the $y$ coordinate $y\leftrightarrow -y$, as motivated by orbifold compactifications. We do not impose any compactness property of the fifth dimension but work in the context of warped geometry so that the background metric $g$ is written as
\begin{equation}
ds^2=g_{MN}dx^Mdx^N=a^2(y) \gamma_{\mu\nu} dx^{\mu}dx^{\nu} + dy^2, \label{bkgdmetric}
\end{equation}
where $\gamma_{\mu\nu}dx^{\mu}dx^{\nu}$ is a $4d$ maximally-symmetric metric, either Minkowski, de Sitter, or anti-de Sitter. This parametrization shows clearly that we look for factorizable solutions, with warp factor $a(y)$ independent of time, and a maximally symmetric brane geometry. Capital Latin letters $M, N$ are used as five-dimensional indices while Greek ones $\mu, \nu$ are used for the indices of the first four dimensions.

We are interested in models containing a k-essence bulk scalar depending only on the fifth coordinate at the level of the background, denoted hereafter by $\phi(y)$. We thus consider the following bulk action, coupling minimally the k-essence bulk field to gravity 
\begin{equation}
S_{bk}=\int d^5x \sqrt{-g}\left\{ \frac R {2\kappa_5^2}  + P(X,\phi) \right\}. \label{5daction}
\end{equation}
The scalar Lagrangian $P$ is an arbitrary function of the scalar field and the standard kinetic term $X$ defined by 
\begin{equation}
X=-\frac 12 g^{MN} \partial_M \phi \partial_N \phi=-\frac 12{\phi'}^2. \label{defX}
\end{equation}
Here and in the following, a prime will denote a derivative with respect to the fifth direction $y$, so that for instance $\phi'=\partial_y\phi$.

In addition to the bulk action, we also consider the brane action, taking the following familiar form  
\begin{align}
S_{bn}&= -\int d^5x  \sqrt{-g_4} V_{bn}(\phi) \delta(y)\nonumber\\
&=-\int d^4x  \sqrt{-g_4} V_{bn}(\phi_0), \label{braneaction}
\end{align}
where $\phi_0=\phi(y=0)$ is the value of the bulk scalar at the position of the brane and $g_4$ is the induced brane four-dimensional metric, $i.e.$ $g_{4\mu\nu}=a^2(y)\gamma_{\mu\nu}$ for the metric \labelcref{bkgdmetric} when the brane is located at $y=0$. 

We make use of the perfect fluid analogy for the stress-energy tensor. Namely, while a perfect fluid respects the isotropy in the rest frame of the fluid elements, the stress-energy tensor of the bulk scalar field here respects the $4d$ maximal symmetry. We then express the stress-energy tensor of the bulk scalar field as
\begin{equation}
T_{MN}=-p_5 g_{MN} + (\rho_5 + p_5) u_Mu_N, \label{SETensor}
\end{equation}
with the five-dimensional `pressure' $p_5$ and `density' $\rho_5$ defined through
\begin{align}
p_5&=-P(X,\phi),  \label{prho1}\\
\rho_5&=P(X,\phi)-2XP_X(X,\phi). \label{prho2}
\end{align}
As in the usual convention, subscripts denote derivatives, so we shall for instance read $P_X=\partial P/ \partial X$.
The 5-velocity $u^M$ has the usual k-essence form, derived from the stress-energy tensor computation. From \cref{defX}, we see that in the case of a bulk k-essence scalar it takes the following simple form
\begin{equation}
u^M=\frac{\partial^M \phi}{\sqrt{-2X}}=(0,0,0,0,1). \label{u}
\end{equation}

Before moving to the bulk equations derived in this setup, we emphasize that the energy `density' and `pressure' introduced above, defined with respect to the bulk direction $y$, are different from the corresponding physical quantities, defined with respect to the time direction. In \cref{physicalrhop} we will introduce these physical quantities to discuss standard energy conditions.

\section{Bulk equations and junction conditions}\label{section:bckgrdandjunctions}

\subsection{Bulk equations}

From the background metric \eqref{bkgdmetric}, one obtains the following non-vanishing Christoffel symbols, components of the Ricci tensor and Ricci scalar
\begin{align}
&\Gamma^y_{\mu\nu}=-a'a\, \gamma_{\mu\nu},  \quad \Gamma^{\mu}_{\nu y}=\frac{a'}{a} \delta^{\mu}_{\nu},  \quad \Gamma^{\rho}_{\mu \nu}=\Gamma^{{\scriptscriptstyle (\gamma)}\rho}_{\hphantom{\scriptscriptstyle (\gamma)}\mu \nu},\nonumber \\
&R_{yy}=-4\frac{a''}a,  \quad  \hspace{10pt} R_{\mu\nu}=R^{\scriptscriptstyle{(\gamma)}}_{\hphantom{\scriptscriptstyle(\gamma)}\!\mu\nu}-(3{a'}^2+a''a)\gamma_{\mu\nu}, \nonumber\\  
&R=\frac{R^{\scriptscriptstyle{(\gamma)}}}{a^2}-12 \frac{{a'}^2}{a^2}-8\frac{a''}{a}.  \label{Gammas}
\end{align}
In the above, the $(\gamma)$ superscripts indicate that the tensors are computed using the four-dimensional metric tensor. In particular, in the maximally symmetric cases under consideration, the four-dimensional Ricci tensor and Ricci scalar read
\begin{equation}
R^{\scriptscriptstyle{(\gamma)}}_{\hphantom{\scriptscriptstyle(\gamma)}\!\mu\nu}=\pm 3h^2\gamma_{\mu\nu}, \qquad R^{\scriptscriptstyle{(\gamma)}}=\pm 12h^2,
\end{equation}
where $h^{-1}$ is the curvature radius of the brane, with thus $h=0$ in the Minkowski case. The signs in the above expressions are positive (negative) for a de Sitter (anti-de Sitter) brane.
The only non-vanishing components of the background Einstein tensor then read 
\begin{align}
&G_{yy}=6\frac{{a'}^2}{a^2}-\frac{ R^{\scriptscriptstyle{(\gamma)}}}{2a^2}, \\
&G_{\mu\nu}=3({a'}^2+a''a)\gamma_{\mu\nu}-\frac14 {R^{\scriptscriptstyle{(\gamma)}}}\gamma_{\mu\nu},
\end{align}
so that the background equations are
\begin{align}
{H}^2&\equiv\left(\frac{a'}{a}\right)^2=\frac{\kappa_5^2}{6} \rho_5 \pm \frac{h^2}{a^2} , \label{Feq1}\\
H'+H^2&=\frac{a''}{a}=-\frac{\kappa_5^2}{6}(2p_5+\rho_5).\label{Feq2}
\end{align}
The stress-energy tensor conservation equation $\nabla_AT^{AB}=0$ leads to the energy conservation equation, which can also be obtained from \cref{Feq1,Feq2},
\begin{equation}
 \rho_5'+4H(p_5+\rho_5)=0.  \label{energycons}
 \end{equation}
In the bulk, the above equation can be written in terms of the k-essence variables $\phi$, $X$, $P(X,\phi)$ as
 \begin{equation}
 \left(P_X+2XP_{XX}\right)\phi''+4{H} P_X \phi' + (P-2XP_X)_{\phi} = 0. \label{energycons2}
 \end{equation}

 \subsection{Junction conditions}

\subsubsection{Israel junction conditions}

The warp factor $a(y)$ is continuous across the brane but its first derivative is not necessarily continuous. Its jump is related to the extrinsic curvature of the brane $K_{\mu\nu}$ and can be expressed, through the Israel junction conditions \cite{Israel:1966rt}, as a function of the (localized) brane stress-energy tensor $S_{\mu\nu}$ and its trace $S$. It reads:
\begin{equation}
[K_{\mu\nu}]=-\kappa_5^2(S_{\mu\nu}-\frac 13 g_{\mu\nu}S). \label{Israel}
\end{equation}
The $[\,\cdot\,]$ notation denotes the jump of any function $f$ across the brane, namely the difference $[f]=f(0^+)-f(0^-)$. We recall that $y=0$ is the localization of the brane. The extrinsic curvature  $K_{\mu\nu}$ is defined as $K_{\mu\nu}=g_{\mu A}g_{\nu B} g^{AC}g^{BD}\nabla_C n_D$ where $n_D$ is the normal vector to the brane. In our case $n=(0,0,0,0,1)$ so that, after expressing the covariant derivative through eqs. \eqref{Gammas}, the extrinsic curvature simply reads
\begin{equation}
K_{\mu\nu}=a'a \gamma_{\mu\nu}.
\end{equation}
The brane stress-energy tensor $S_{\mu\nu}$ can be expressed simply from the brane action \eqref{braneaction}. It takes the form 
\begin{equation}
S_{\mu\nu}=-g_{\mu\nu}V_{bn}=-a^2(0) \eta_{\mu\nu}V_{bn},\label{surfaceSET}
\end{equation}
 so that \cref{Israel}  leads to
\begin{equation}
\frac{[a']}{a(0)}=-\frac{\kappa_5^2}{3}V_{bn}(\phi_0).
\end{equation}
We also recall that $\phi_0=\phi(y=0)$. The value of $V_{bn}(\phi_0)$ is thus naturally related to the brane tension.

The $\mathbb{Z}_2$ symmetry along $y$ implies that $[a']=2a'(0^+)$, so that the Israel junction conditions require 
\begin{align}
H(0^+)=\frac{a'}{a}(0^+)&=-\frac{\kappa_5^2}{6}V_{bn}(\phi_0) \label{Vphi0}\\
&=-{\rm sgn}(V_{bn})\sqrt{\frac{\kappa_5^2}{6}\rho_5(0^+)\pm \frac{h^2}{a^2(0)}}.\nonumber
\end{align}
We used the Friedmann equation \labelcref{Feq1} to obtain the last equality expressed in terms of $\rho_5$.

\subsubsection{Scalar field junction conditions}\label{subsec:scalarjunctionconditions}

 As for the first derivative of the warp factor, there might exist a jump in the first derivative of the k-essence scalar. Indeed, due to the
$\mathbb Z_2$ symmetry across the brane (localized at $y=0$) we must have $\phi'(-y)=-\phi'(y)$. This discontinuity is extracted by taking into account the brane potential in the scalar field equation of motion, $i.e.$ by rewriting \cref{energycons} as
\begin{equation}
\frac 1{a^4}\partial_y\left(a^4 P_X\phi'\right) + P_{\phi} - V_{{bn}, \, \phi}(\phi) \delta(y)=0, \label{fieldequationscalar}
\end{equation}
and integrating across the brane, namely keeping only the $\delta(y)$ dependent parts, to obtain
\begin{equation}
\frac 1{a^4}\left[a^4 P_X \phi'\right]=2P_X \phi'|_{0^+}=V_{{bn}, \, \phi}(\phi_0). \label{Vpphi0}
\end{equation}
This relation shows that the jump of the derivative of the bulk scalar is supported by the derivative of the brane scalar potential.

\subsubsection{Key equations}

Using \cref{prho1,prho2}, the squares of the junction conditions \labelcref{Vphi0,Vpphi0} can be written in terms of  $P$ and  $X$ as 
\begin{align}
-8XP_X^2|_{0^+}&=V_{bn,\,\phi}^2(\phi_0), \label{junction2}\\
(P-2XP_X)|_{0^+}\pm \frac{6 h^2}{\kappa_5^2 a^2(0)}&=\frac {\kappa_5^2}6 V_{bn}^2(\phi_0). \label{junction1}
\end{align}

\section{Requirements for a self-tuning cosmological constant} \label{section:selftuning}

We now review the logic of the self-tuning mechanism of the cosmological constant introduced in \cite{Arkani-Hamed:2000hpr,Csaki:2000wz,Kachru:2000hf} and extend it to the case of a bulk k-essence scalar. 

For the present discussion, we suppose that the bulk scalar coupling to the matter fields is of the Brans-Dicke type \cite{Brans:1961sx}, realized only through the Jordan frame metric 
\begin{equation}
 \tilde{g}_{4 \mu \nu}=f^{\frac 12}(\phi)g_{4 \mu\nu},
\end{equation}
where $f(\phi)$ is the effective coupling of the bulk scalar to the matter fields on the brane, so that there is no violation of the weak equivalence principle. This setup is motivated not only by the stringent experimental constraints on the violation of the weak equivalence principle but also by the fact that such a universal coupling is unaffected by loop corrections from the brane fields \cite{Arkani-Hamed:2000hpr}. (We come back shortly to the different frame descriptions of our model in \cref{subsection:equivalentactions}.) From \cref{scalarpotdecomp,braneaction}, this form of the universal matter coupling implies that the effective vacuum energy $V_0$ from the brane sector is related to the brane scalar potential $V_{bn}$ as 
\begin{equation}
V_{{bn}}(\phi)=V_0 f(\phi). \label{scalarpotdecomp}
\end{equation}

\subsection{Requirement from junction conditions}\label{requirementfromjunction}

The junction conditions \labelcref{junction2,junction1} then read
\begin{align}
-8XP_X^2|_{0^+}&=V_0^2 f_{\phi}(\phi_0)^2, \label{junction2bis}\\
(P-2XP_X)|_{0^+}\pm\frac{6h^2}{\kappa_5^2a^2(0)}&=\frac {\kappa_5^2}6 V_0^2 f(\phi_0)^2. \label{junction1bis}
\end{align}

If the functions $P(X,\phi)$ and $f(\phi)$ are such that the conditions \labelcref{junction1bis,junction2bis} enforce $h=0$ independently of $V_0$, the brane cosmological constant is said to be self-tuned to a vanishing value. 
Said differently, we talk about self-tuning of the $4d$ cosmological constant, or self-tuning mechanism in short, for models where the flat brane solution $h=0$ is {\it enforced} (at least for maximally symmetric brane geometries) through the junction conditions, based on the shape of the brane scalar potential and of the bulk components, independently of the value of the brane effective vacuum energy $V_0$.

In the original literature, `self-tuning' was sometimes used in a looser way, as soon as flat brane solutions existed for any vacuum energy $V_0$, even if not preferred with respect to $h\neq0$ solutions. We do not retain this meaning, because we do not consider that self-tuning is achieved in the cases where the flat brane solution is not preferred.

Since we consider a general function $P(X,\phi)$ for the bulk action, one can always redefine $\phi$ so that the effective universal coupling $f(\phi)$ of the bulk scalar to the matter fields is of the form
\begin{equation}
 f(\phi)=e^{\gamma\phi}, \label{branescalarpot}
\end{equation}
where $\gamma$ is an arbitrary constant. This form of $f(\phi)$ was considered in \cite{Arkani-Hamed:2000hpr} and also simplifies the analysis below. An important difference between the treatment in \cite{Arkani-Hamed:2000hpr} and here is that, while the choice of $f(\phi)$ in \cite{Arkani-Hamed:2000hpr} is a physical assumption (since the bulk action there is fixed), here the choice of $f(\phi)$ simply fixes the definition of $\phi$ (since $P$ is a priori a general function of $X$ and $\phi$). 

With (\ref{branescalarpot}), we have $V_{bn, \phi}(\phi)=\gamma V_{bn}(\phi)$ so that \eqref{junction2bis} and \eqref{junction1bis} read 
\begin{align}
-8XP_X^2|_{0^+}&=\gamma^2 V_{{ bn}}^2(\phi_0), \label{junction2-example}\\
(P-2XP_X)|_{0^+}\pm\frac{6h^2}{\kappa_5^2a^2(0)}&=\frac {\kappa_5^2}6 V_{ bn}^2(\phi_0).\label{junction1-example}
\end{align}
Hence, we see that if $P(X)$ is chosen such that 
\begin{equation}
\frac {4\kappa_5^2}{3\gamma^2}XP_X^2-2XP_X+P =0, \label{ode-for-P(X)}
\end{equation}
the set of junctions conditions \labelcref{junction1-example,junction2-example} enforces $h=0$. As  equation \labelcref{ode-for-P(X)} does not depend on $V_0$, we are exactly in the case of  $4d$ cosmological constant self-tuning.

\Cref{ode-for-P(X)} can be rewritten as
\begin{equation}
 P_x^2 - 2P_x + \frac{P}{x} = 0, \qquad {\rm with} \quad x\equiv \frac{3\gamma^2}{4\kappa_5^2}X, \label{eqn:ode-for-P(x)}
\end{equation}
and can be considered as a partial differential equation for $P(x,\phi)$. This equation leads to $P_x=\pm \sqrt{1-P/x}+1$, which is of the homogeneous type. It is solved by defining $u$, through $P=ux$, which satisfies the separable differential equation
\begin{equation}
\frac{d u}{d x} x+u = \pm \sqrt{1-u}+1. \label{ode-u}
\end{equation}
There are two types of solutions:
\begin{enumerate}
\item If $u$ is constant, \cref{ode-u} is solved by $u=1$ (or $u=0$), which simply gives 
\begin{equation}
P=x=\frac{3\gamma^2}{4\kappa_5^2}X.
\end{equation}
This is exactly the case studied in \cite{Arkani-Hamed:2000hpr}, with $\gamma=2\kappa_5$. The solution for $P$ gives indeed their kinetic term $-{\frac 32} \, {\phi'}^2$ for the bulk scalar. We recall that this solution suffers from bulk singularities at finite distances.
\item When $u$ is not constant, one can again change variables. Let us define $z=\sqrt{1-u}$, such that $du=-2zdz$. Equation \eqref{ode-u} then leads to the separable equation $2 x{dz}=-(z \pm 1)dx$, solved by the form $z=[x/c(\phi) ]^{-1/2}\pm1$ leading to
\begin{equation}
 P = \pm 2\sqrt{c(\phi)x} - c(\phi),  \quad x= \frac{3\gamma^2}{4\kappa_5^2}X,\label{sqrtselftuningsolution}
\end{equation}
where $c(\phi)$ is an arbitrary function of $\phi$ and is negative because in our case  $x<0$.
This  solution describes a Cuscuton field with a negative bulk potential $c(\phi)$. It thus gives a similar scenario as in the original Randall-Sundrum models \cite{Randall:1999vf,Randall:1999ee}, with the addition of a Cuscuton field and the promotion of the cosmological constant to a potential. The kinetic term of this specific form does not contribute to the bulk energy density but plays an important role to achieve the self-tuning mechanism. We study this solution in more detail in the following.
\end{enumerate}

Note that the above solutions are global solutions. It might be possible to impose these solutions near the brane while changing the shape of $P$ away from the brane, $i.e.$ after some (single or multiple) critical values for $X$. Nevertheless this would necessarily imply some engineering since the bulk evolution of $X$, namely the dependence $X(y)$ in the fifth coordinate, is only determined after resolution of the field equations. We have not further explored this possibility.

\subsection{Requirement from bulk equations} \label{section:Cuscuton}

We now study in more detail the self-tuning mechanism realized through the solution given in \eqref{sqrtselftuningsolution}, which can be interpreted as a bulk Cuscuton field \cite{Afshordi:2006ad}. As described in \cref{section:stability}, causality requires choosing the solution with a negative sign in \cref{sqrtselftuningsolution}. We are thus interested in the self-tuning solution corresponding to 
\begin{equation}
 P = -2\sqrt{c(\phi)x} - c(\phi),  \quad x= \frac{3\gamma^2}{4\kappa_5^2}X<0, \quad c(\phi)<0, \label{sqrtselftuningsolution2}
\end{equation}
with the brane scalar potential given in \cref{scalarpotdecomp,branescalarpot}, that we repeat here
\begin{equation}
 V_{bn}(\phi) = V_0 f(\phi)\,, \qquad f(\phi)=e^{\gamma\phi}. \label{recallVbrane}
\end{equation}
The sign of $\gamma$ can be absorbed in the definition of $\phi$, hence we will consider $\gamma>0$  in the following.

The bulk energy density and pressure given by \labelcref{prho1,prho2} read 
\begin{align}
p_5&=\frac{\sqrt{3}\gamma}{\kappa_5} \sqrt{c(\phi)X} + c(\phi), \label{cuscup5}\\ 
 \rho_5&=-c(\phi).  \label{cuscurho5}
\end{align}
The field equations \labelcref{Feq1,energycons} are thus
\begin{align}
&H^2=\frac{\kappa_5^2}{6}\rho_5=-\frac{\kappa_5^2}{6}c(\phi), \label{Feq1cphi}\\
&4HP_X\phi'-c_{\phi}(\phi)=0. \label{Feq2cphi}
\end{align}
This system of equations is inconsistent for constant $c$, corresponding to a Cuscuton with a bulk cosmological constant. Nevertheless, a non-trivial solution is possible when the function $c(\phi)$ satisfies
\begin{align}
c_{\phi}(\phi)&={\rm sgn}(\phi') \frac{2\sqrt{6}\gamma}{\kappa_5} H\sqrt{-c(\phi)}\nonumber\\
&=-{\rm sgn}(\phi'H) \,\, 2\gamma \, c(\phi). \label{equadiffcphi}
\end{align}
The sign in the above equation can be traced back to the brane junction conditions \labelcref{Vphi0,Vpphi0}, which necessitate near the brane
\begin{equation}
\hspace{-0.1cm}{\rm sgn}(\phi'H)=-{\rm sgn}(V_{bn}V_{bn,\,\phi})=-{\rm sgn}(\gamma V_{bn}^2).
\end{equation}
Hence, for our positive brane `charge' $\gamma>0$, we obtain ${\rm sgn}(\phi'H)=-1$. The solution of \cref{equadiffcphi} is thus of the form
\begin{equation}
c(\phi)=-c_0 e^{2\gamma \phi}, \qquad c_0>0. \label{solutionforcphi}
\end{equation}

The action implementing the cosmological constant self-tuning mechanism in presence of a bulk Cuscuton field is thus
\begin{align}
S&=\!\!\int \!\!d^5x \sqrt{-g}\left\{ \frac R{2\kappa_5^2} \! -\! \frac{\sqrt{3}\gamma}{\kappa_5} e^{\gamma \phi} \sqrt{-c_0 X} + c_0 e^{2\gamma\phi} \right\} \nonumber\\
& \hspace{10pt} - \!\!\int \!\! d^5x \sqrt{-g_4} \,\, V_0 e^{\gamma\phi}\delta(y). \label{5dactionfinal}
\end{align}

\subsection{Scalar field and warp factor profiles}
The first Friedmann equation \eqref{Feq1cphi}  simply relates the bulk scalar to the $H$ bulk parameter through 
\begin{equation}
\phi=  \frac 1\gamma \left ( \ln | H | + \frac 12 \ln\left (\frac6{c_0\kappa_5^2}\right)\right). \label{relationphiH}
\end{equation}
Hence any given warp factor $a(y)$, or yet any $H(y)$, can be supported by the bulk solution $\phi(y)$ given through \cref{relationphiH} ! This means that the action \labelcref{5dactionfinal} does not determine the brane profile (nor the scalar field profile). We expect the brane profile to be determined from initial conditions and evolution of the brane fields, once considered in a cosmological context. If stable, our solution should indeed be seen as the final stage of the cosmic evolution of the system, once all the brane fields are diluted by the brane expansion. Before dilution, the brane fields would have determined the warp factor profile through their coupling to the bulk scalar.

We conclude this part by recalling that localization of gravity on the brane requires {in particular} a finite $4d$ Planck mass. The effective $4d$ Planck mass is extracted from the coefficient in front of the effective $4d$ Einstein-Hilbert term for the metric perturbations around our background, after integration along the fifth dimension. It is given by
\begin{equation}
\frac1{\kappa}=\frac{M_P^2}{8\pi}=\frac1{\kappa_5^2}\int dy \, a(y)^2.
\end{equation}
Hence, we see that even if arbitrary warp factor profiles can be supported by our model, localization of gravity constrains $a(y)$ to be a square-integrable function. One usually considers warp factors decreasing when moving away from the brane, thus with $H(y)<0$ for $y>0$. {We eventually emphasize that finiteness of the $4d$ Planck mass is only a necessary condition for localized gravity. One should further require that the graviton massive modes are sufficiently suppressed around the brane. This condition is expected to further restrict warp factor profiles close to the brane, as shall be studied by solving the metric perturbations equations of motion along the fifth dimension with the proper boundary conditions near the brane and at infinity. Furthermore, due to the presence of the bulk scalar, the role of bending modes may be more intricate than in the original Randall-Sundrum scenario \cite{Randall:1999vf,Garriga:1999yh,Karch:2000ct,Giddings:2000mu}.}

\subsection{Equivalent actions}\label{subsection:equivalentactions}
For completeness, we expose hereafter different formulations of our model, obtained either by rescaling the metric or redefining the bulk scalar.
 
\subsubsection{Other frames}
Under a local Weyl rescaling  depending on the bulk scalar, $i.e.$ after defining a new metric $\bar g$ through $g_{MN}=\exp(-\omega \phi) \bar g_{MN}$ for arbitrary parameter $\omega$,  our action \eqref{5dactionfinal} is expressed as
\begin{align}
\hspace{0cm}S&=\int d^5x \sqrt{-\bar g}\left\{ \frac{e^{-\frac32 \omega\phi}}{2\kappa_5^2}\left({\bar R}  - 6\omega^2 \bar X\right)  \right. \nonumber\\
&\hspace{0.7cm}- \left. \frac{\sqrt{3}\gamma}{\kappa_5} e^{(\gamma -2\omega)\phi} \sqrt{-c_0 \bar X} + \vphantom{\frac12} c_0 e^{(2\gamma-\frac 52\omega) \phi}\right\}  \nonumber \\
& \hspace{0.7cm} - \int d^5x \sqrt{-\bar{g}_4}\, V_0 e^{(\gamma-2 \omega)\phi}\delta(y).   \label{Weylrescaledaction}
\end{align}

The Jordan frame, for which the brane field and bulk scalar couple only through the metric $\bar g_{MN}$, is thus reached by a Weyl rescaling with  parameter $\omega=\gamma/2$. In this frame, our model is described by the following action
\begin{align}
\hspace{-0.5cm}S_J=\int d^5x \sqrt{-\bar g}&\left\{ e^{-\frac34 \gamma\phi}\left(\frac{\bar R}{2\kappa_5^2}  +c_0\right)  \right. \nonumber\\
& \,\,\,- \left. \frac 34 \frac{\gamma^2}{\kappa_5^2} \bar X - \frac{\sqrt{3}\gamma}{\kappa_5}  \sqrt{-c_0 \bar X} \right\} \hspace{-0.5cm}  \nonumber \\
&\hspace{-1cm }-\int d^5x \sqrt{-\bar g_4}\, V_0\delta(y). \label{5dactionfinalJordan}
\end{align}

We see from \cref{Weylrescaledaction} that after a Weyl rescaling with parameter $\omega=2\gamma$, the action \eqref{5dactionfinal} transforms to the `string frame' action
\begin{align}
\hspace{0cm}S_{\rm s. f.}&=\int d^5x \sqrt{-\bar g} e^{-3\gamma \phi} \left\{ \frac{\bar R}{2\kappa_5^2}  +  c_0  \right. \nonumber\\
&\hspace{2cm}-  \left. 12 \frac{\gamma^2}{\kappa_5^2} \bar X  - \sqrt{3}\frac{\gamma}{\kappa_5} \sqrt{-c_0 \bar X} \right\} \nonumber \\
&\hspace{1cm }- \int d^5x e^{-3\gamma \phi} \sqrt{-\bar g_4}\, V_0 \delta(y) .
\end{align}
In this frame, the bulk scalar couples each part of the action in the exact same way: the Einstein-Hilbert term, bulk cosmological constant $c_0$ and brane vacuum energy $V_0$ (or tension) all scale as $\exp(-3\gamma\phi)$, as would be the case for the string dilaton (but with a standard kinetic term). This fact already appeared in \cite{Kachru:2000hf}, in the discussion under their $(2.3)$. Their action $(2.1)$ is  similar to the one we are studying, except for the crucial difference that in our case we consider a Cuscuton bulk field, namely a peculiar kinetic term, allowing for self-tuning without singularity.

\subsubsection{Pure Cuscuton kinetic term} 
If one likes to stick to a pure Cuscuton kinetic term (with no $\phi$ dependence in the square root), one can also define a new variable $\tilde \phi$ such that $c(\phi)X=-\tilde X$, namely such that $d\tilde \phi=\sqrt{-c(\phi)}d\phi=\sqrt{c_0} e^{\gamma \phi} d\phi$ with our particular choice of $c(\phi)$ given in \cref{solutionforcphi}. This leads to the relation
\begin{equation}
\tilde\phi=\frac{\sqrt{c_0}}{\gamma}e^{\gamma\phi}, \label{tildephi}
\end{equation}
so that the model defined through \cref{sqrtselftuningsolution2} is equivalent to the one for $\tilde \phi $ with
\begin{align}
&P(\tilde X,\tilde \phi)=-\frac{\sqrt{3}\gamma}{\kappa_5} \sqrt{-\tilde X}+{\gamma^2}\tilde\phi^2, \\
  &\tilde V_{bn}(\tilde \phi)=V_{bn}(\phi)=V_0e^{\gamma\phi}=\frac{\gamma V_0}{\sqrt{c_0}}\tilde\phi.
\end{align}
Our action \labelcref{5dactionfinal} is thus equivalent to the action 
\begin{align}
S&=\!\!\int \!\!d^5x \sqrt{-g}\left\{ \frac R{2\kappa_5^2} \! -\! \frac{\sqrt{3}\gamma}{\kappa_5}\sqrt{-\tilde X} +{\gamma^2}\tilde\phi^2 \right\} \nonumber\\
& \hspace{10pt} - \!\!\int \!\! d^5x \sqrt{-g_4} \,\,\frac{\gamma V_0}{\sqrt{c_0}}\tilde\phi \, \delta(y), \label{5dactionfinaltilde}
\end{align}
for which the kinetic term is the standard Cuscuton one, depending on $\tilde X$ only, associated to a quadratic bulk scalar potential for $\vphantom{3^{3^3}} \tilde \phi$.

\section{Causality and stability of the self-tuning solution with Cuscuton bulk field} \label{section:stability}

\subsubsection{Infinite sound speed and causality}\label{soundspeedcausality}The bulk Cuscuton `sound speed' is defined as usual for k-essence fields
\begin{equation}
c_5^2=\frac{P_X}{P_X+2XP_{XX}}=\infty. \label{cs}
\end{equation}
The Cuscuton sound speed is infinite by construction, since the denominator in \cref{cs} vanishes. This immediately raises the question of causality in such models, but as described in the original study of the Cuscuton solution \cite{Afshordi:2006ad}, a careful analysis shows that there is no problem, in the sense of the propagation of local signals.

 Following the argument of the original study \cite{Afshordi:2006ad},  in the space-like case $X<0$ considered here, one should nevertheless impose a negative sign in front of the square root in the action. This fact is equivalent to requiring the correct sign for the kinetic term of the scalar field perturbations around our considered background, hence related to the absence of ghost.
 
Hence, even if the sign in front of the square root of the self-tuning solution \eqref{sqrtselftuningsolution} is {\it a priori} arbitrary, causality constrains it to be negative. This justifies the choice made at the beginning of \cref{section:Cuscuton} to consider only the solution \eqref{sqrtselftuningsolution2}.

\subsubsection{Stability} \label{stabilitysection} The precise study of the stability of our self-tuning background solution with bulk Cuscuton is beyond the scope of the present paper. We nevertheless make some comments and draw parallels with what is known for standard $4d$ Cuscuton fields. 

First of all, the scalar field perturbations are simply related to the scalar metric perturbations, due to the specific Cuscuton shape of our bulk k-essence scalar. We introduce the field scalar field $\tilde \phi_p$, perturbed around the background solution $\tilde \phi(y)$ described previously. For simplicity of the formulae, we consider the field $\tilde \phi$ introduced in \cref{tildephi} rather than $\phi$, but omit the tilde hereafter. We expand the perturbation at linear order as
\begin{equation}
\phi_p(x^{\mu},y)=\phi(y)+\delta\phi(x^\mu,y).
\end{equation}
We also introduce scalar metric perturbations in longitudinal gauge. This gauge cannot be fully fixed if we want to keep the position of the brane at $y=0$, but here we just comment on generic bulk perturbations and forget for now the presence of the brane. In this gauge, one obtains the relation
\begin{equation}
\delta \phi =-{\rm sgn}(\phi')\frac{\sqrt{6}}{\kappa_5 \gamma}\left( \Phi'+2H\Phi \right),
\end{equation}
with $\Phi$ similar to the usual Bardeen parameter, thus related to the metric scalar perturbations (in particular warp factor perturbations.) As before, the prime in $\Phi'$ denotes derivative along the fifth dimension. As advertised, we see that there is a simple relation between the scalar field perturbation and the scalar metric ones, as is also the case in  Cuscuton cosmology \cite{Afshordi:2007yx}. Here, as the bulk parameter $H(y)$ is only supported by the Cuscuton scalar,  the relation between scalar field and metric perturbations is even simpler and depends only on the fifth dimension scales, through $\Phi'$.

The field equations result in the simple wave equation for the Bardeen parameter
\begin{equation}
\Box \Phi=0, \label{waveeqbulkperturb}
\end{equation}
where $\Box$ denotes the $4d$ d'Alembertian. 

This equation can be directly inferred by analogy to the dispersion relation for perturbations of a perfect fluid in $4d$ cosmology, which reads
\begin{equation}
\omega^2=c_s^2 \frac{\bold{k}^2}{a^2} - \frac{\kappa \rho}{2}. \label{dustdispersion}
\end{equation}
The analogy is worked out by exchanging the cosmological time with the brane-world fifth dimension and the cosmological spatial coordinates with our first four coordinates. The dispersion relation for perturbations in the braneworld setup can thus be inferred from the dust perturbation dispersion relation in cosmology by exchanging $\omega^2\leftrightarrow -k_y^2$, $\bold{k}^2\leftrightarrow k^{\mu}k_{\mu}$ and  $c_s\leftrightarrow c_5$. It takes the form 
\begin{equation}
-k_y^2=c_5^2 \frac{k^{\mu}k_{\mu}}{a^2} - \# {\kappa_5^2 \rho_5}. \label{cuscutonpertdispersion}
\end{equation}
The $\#$ symbol represents a numerical coefficient expected to be different in the analog dispersion relation.
As mentioned in \cref{cs}, the $5d$ `sound speed'  $c_5$ for a Cuscuton field is infinite. In this limit, the analog dispersion relation \eqref{cuscutonpertdispersion} reduces to
\begin{equation}
k^{\mu}k_{\mu}=0.
\end{equation}
This is exactly the dispersion relation for the modes of perturbations satisfying the wave equation \eqref{waveeqbulkperturb}, which respects the $4d$ Lorentz invariance parallel to the world-volume of the brane. Furthermore, \eqref{cuscutonpertdispersion} clearly shows that the physical sound speed $c_s^{\perp}$ in the direction perpendicular to the brane, $i.e.$ the $y$ direction, is related to the $5d$ `sound speed'  $c_5$ as $c_s^{\perp}=1/c_5$. Hence equation \eqref{cs} simply states that $c_s^{\perp}=0$ and shows that, for our static background solution, perturbations do not propagate in this direction.

In fact, \cref{waveeqbulkperturb} reveals that scalar perturbations are just waves propagating freely on surfaces orthogonal to the $y$ direction. Again, this is analog to what was described originally for perturbations in a $4d$ Cuscuton background with $X<0$, which behave like waves propagating on constant mean curvature surfaces \cite{Afshordi:2006ad}. In our case, this shows that there is no obvious instability for the bulk scalar perturbations introduced far from the brane. A full stability analysis should also take into account the presence of the brane and eventual stability issues coming from the perturbed junction conditions once the brane bending modes are taken into account \cite{Garriga:1999yh,Karch:2000ct,Giddings:2000mu,Mukohyama:2000ga,Mukohyama:2000ui}. This is left for future work.

\subsubsection{Standard energy conditions for physical energy density and pressure}\label{physicalrhop}
As mentioned at the end of \cref{section:kessencebulk}, one can introduce the physical energy density and stress of our setup, interpreted as an anisotropic fluid in five dimension. The physical quantities can indeed be related to the five-dimensional ones $\rho_5$ and $p_5$ by rewriting the stress-energy tensor of \cref{SETensor} in the following way:
\begin{align}
T_{MN}&=\rho_5\,  u_Mu_N - p_5 (g_{MN} - u_Mu_N) \nonumber \\
&= \rho \, v_Mv_N+ p^{\parallel} (g_{MN}-u_Mu_N+v_Mv_N) \nonumber \\
&\,\,\,\, +p^{\perp} u_Mu_N \label{SETensorPhysical}
\end{align}
We introduced the physical energy density $\rho$, the pressures $p^{\parallel}$ and $p^{\perp}$ respectively parallel and orthogonal to the brane,  as well as the 5-velocity of the physical fluid $v_M=(-\sqrt{|g_{00}|},0,0,0,0)$, normalized with respect to the background metric of \cref{bkgdmetric}. The 5-velocity $u_M$ was defined in \cref{u}.

Note that by introducing the diagonal stress tensor 
\begin{equation}
\tau_{M}^{\hphantom{M}N}\equiv \text{Diag}(0,p^{\parallel},p^{\parallel},p^{\parallel},p^{\perp}),
\end{equation}
one can re-express the stress-energy tensor generically as
\begin{equation}
T_{MN}= \rho v_Mv_N + \tau_{MN}. \label{STensor}
\end{equation}
The stress tensor can be decomposed into the physical isotropic pressure $p$ and anisotropic stress $\pi_{MN}$ as
\begin{equation}
 \tau_{MN}=p\, (g_{MN}+v_Mv_N) + \pi_{MN},
 \end{equation}
with
\begin{align}
&p=\frac{1}{4} \text{Tr}(\tau_{MN})=\frac{3}{4} p^{\parallel} + \frac{1}{4} p^{\perp}, \\
 &\pi_{M}^{\hphantom{M}N}=\frac{p^{\parallel}-p^{\perp}}{4} \text{Diag}(0,1,1,1,-3).
\end{align}

By identifying (in components) the first line of \cref{SETensorPhysical} with its last line or with \cref{STensor}, one can then easily relate the physical quantities of the anisotropic fluid to the bulk quantities. Indeed, the diagonal component of the stress-energy tensor read
\begin{align}
&T_{00}= -p_5 g_{00} = (\rho + p^{\parallel})|g_{00}|+p^{\parallel}g_{00}=-\rho g_{00} \nonumber \\
&T_{11}= -p_5 g_{11}= p^{\parallel} g_{11} \\
&T_{yy}= \rho_5=p^{\perp}, \nonumber
\end{align}
so that we can identify
\begin{equation}
\rho=p_5, \qquad p^{\parallel}=-p_5, \qquad p^{\perp}=\rho_5.
\end{equation}

We now turn to the study of energy conditions. One could derive them directly from the stress-energy tensor expressed in terms of (unphysical) bulk `density' and `pressure' $\rho_5$ and $p_5$. Nevertheless, they are usually expressed and interpreted in terms of the standard physical quantities. They read:
\begin{align}
&\text{\it null:}\,  \left\{
    \begin{array}{ll}
        \rho+p^{\parallel}=0\geq 0 \\
         \rho+p^{\perp}=p_5+\rho_5 \geq 0
    \end{array}
\right. \, , \label{ec1}\\
&\text{\it weak:} \, \left\{
    \begin{array}{ll}
        null \\
       \rho=p_5\geq 0
    \end{array}
    \right. \,, \label{ec2}\\
&\text{\it strong:} \,  \left\{
    \begin{array}{ll}
        null \\
      \rho + 3 p^{\parallel} + p^{\perp}=-2p_5+\rho_5 \geq 0
    \end{array}
    \right. \,. \label{ec3} 
\end{align}
The weak and strong energy conditions can be easily violated in presence of scalar fields. For instance, a single scalar field with negative potential and small kinetic energy violates the weak energy condition whereas an inflationary phase violates the strong one. We thus do not expect them to be related to stability of the system in our setup, which describes a bulk scalar.

On the other hand, the null energy condition is often related to stability and/or causality of perturbations around the solution, in particular the absence of ghost. We thus check that the null energy condition is satisfied for the self-tuning solution obtained previously. The five-dimensional `pressure' and `density' for the solution were given in \cref{cuscup5,cuscurho5}, so that the null energy condition \eqref{ec1} reads
\begin{equation}
null: \rho_5+p_5= \frac{\sqrt{3}\gamma}{\kappa_5} \sqrt{c(\phi)X} \geq 0, \label{nullself}
\end{equation}
which is always satisfied. Indeed, we recall that we chose $\gamma>0$, see the discussion at the beginning of \cref{section:Cuscuton}. The condition \cref{nullself} is satisfied precisely due to the choice of (negative) sign in front of the square root in the solution \eqref{sqrtselftuningsolution}. This sign was fixed at the very beginning of \cref{section:Cuscuton}, precisely motivated by causality arguments, as explained later in \cref{soundspeedcausality}. 

The first part of the null energy condition  \eqref{ec1}, which is trivially satisfied here, is simply a consequence of the $4d$ Lorentz invariance of the solution. On the other hand, by inspecting the quadratic action for bulk perturbations it is seen that \eqref{nullself} is equivalent to the non-negativity of the time kinetic term of the perturbations, $i.e.$ to the no-ghost condition.

To conclude, we have just seen that the null energy condition, which holds in the present setup, nicely reflects the stability of our k-essence model and its self-tuning solution.

\section{Summary and discussion}\label{section:discussion}

In this work, we developed a brane-world model achieving self-tuning of the cosmological constant. Junction conditions enforce the brane to stay flat, independently of its effective vacuum energy, as long as maximally symmetric solutions are considered. We investigated brane-world scenarios with k-essence bulk scalars and showed that imposing the self-tuning mechanism greatly constrains the possible Lagrangians. Our model thus employs a specific kinetic term for the bulk scalar. The latter can be identified as a spacelike Cuscuton after the brane scalar potential is brought to an exponential form. Our solution is very close to the Lagrangian studied previously in \cite{Forste:2011hq}, but these authors explicitly ignored the choice of parameter leading to our model. 

We found that, at the background level, the Cuscuton bulk scalar can accommodate any profile for the metric warp factor. In particular, it can describe warp factor profiles without bulk singularity and with finite four-dimensional Planck mass, which is a necessary condition for gravity to be localized on the brane. 

A quick view of the bulk scalar perturbations motivated our belief that our solution is not trivially unstable. Nevertheless, a precise stability analysis should incorporate the presence of the brane, in particular by inspecting the bending modes perturbations and their influence on the brane junction conditions.

Our solution being static, it can at best describe the state of our Universe at late times. A complete picture must describe the cosmic history of the brane, in the spirit of cosmological brane-world solutions developed initially in \cite{Binetruy:1999ut,Binetruy:1999hy,Mukohyama:1999qx}. It would thus be interesting to analyze how an expanding brane Universe would interact with the Cuscuton bulk scalar and under which conditions it could lead to our static solution, after dilution of the brane fields by cosmic expansion at late times. We expect that the cosmic history of the brane finally determines the warp factor profile of our static solution.

The (timelike) Cuscuton theory in the context of cosmology is closely related to the VCDM modified gravity theory \cite{DeFelice:2020eju,DeFelice:2020cpt}. Similarities and differences between these two theories have been studied in \cite{Aoki:2021zuy,DeFelice:2022uxv}. It is certainly worthwhile investigating brane-world scenarios based on the five-dimensional and spacelike version of the latter theory and examining if the self-tuning mechanism holds in that case.

Finally, it would be particularly interesting to get deeper insights into the origin of our effective action by studying eventual ultraviolet completions flowing to our peculiar bulk effective action. Progress in that direction could start by embedding our setup in a supersymmetry (or rather supergravity) framework following lines developed in \cite{Khoury:2010gb,Farakos:2012je,Farakos:2012qu,Koehn:2012ar} for models including k-essence and higher-derivative terms.

\section*{Note Added}

In the very final stage of completion of the present paper, works sharing ideas with our study were brought to our attention \cite{Andrade:2018afh,Bazeia_2021}. These works also make use of a Cuscuton term, but this last one is added on top of standard kinetic terms. Moreover, their authors do not study self-tuning mechanisms.

\section*{Acknowlegments}

The work of OL is supported in part by Japan Society for the Promotion of Science Grant-in-Aid for Scientific Research No. 17H06359. 
The work of SM is supported in part by Japan Society for the Promotion of Science Grants-in-Aid for Scientific Research No. 17H02890. , and No. 17H06359 and by World Premier International Research Center Initiative, The Ministry of Education, Culture, Sports, Science and Technology, Japan.

\appendix

\section{Equation of state and warp factor reconstruction}\label{appendix:A}

\paragraph{Equation of state}
 The equation of state for the bulk k-essence scalar directly follows from the definitions \labelcref{prho1,prho2}: 
\begin{equation}
w\equiv\frac{p_5}{\rho_5} = \frac{P(X,\phi)}{2XP_X(X,\phi)-P(X,\phi)}. \label{eos}
\end{equation}
Hence, the choice of function $P(X,\phi)$ determines the bulk equation of state. Reciprocally, one can try and reconstruct any bulk equation of state by solving \cref{eos} regarded as a first-order differential equation for  $P(X,\phi)$. 

For instance, the non-linear fluid equation of state introduced in \cite{Antoniadis:2021rxw} to avoid bulk singularities,
\begin{equation}
p_5=\ell \rho_5^{\lambda}, \label{nonlinearfluideos}
\end{equation} 
can be obtained (for $\lambda\neq1$) by choosing 
\begin{equation}
P(X)= (-\ell)^{\frac{1}{1-\lambda}}\left(1+(-X)^{\frac{\lambda-1}{2\lambda}}\right)^{\frac \lambda{\lambda-1}}. \label{nonlinearPX}
\end{equation}
The case with $\lambda=3/2$, studied analytically in \cite{Antoniadis:2021rxw}, is thus obtained for 
\begin{equation}
P(X)= \frac 1{\ell^2}\left(1+(-X)^{\frac 16}\right)^3. \label{nonlineareos}
\end{equation}

We also comment on the fact that the `unorthodox' Lagrangians studied in \cite{Forste:2011hq} can be recovered from the bulk k-essence formalism by considering the function
\begin{equation}
P(X)=\lambda X^{\alpha}.
\end{equation}
A small difference resides in the fact that this Lagrangian describes a real scalar, whereas \cite{Forste:2011hq} originally considered complex scalars.
{The goal of the main body of the paper is to show that requiring self-tuning in our particular setup enforces a particular equation of state, derived from the k-essence function $P(X,\phi)$ shown in \cref{sqrtselftuningsolution}.}

 \paragraph{Warp factor reconstruction} 
One can theoretically reconstruct a generic warp factor profile $a(y)$ by choosing an appropriate function $P(X)$. The choice is not unique. If one is interested in shift-symmetric cases (with $h=0$ for simplicity), the reconstruction of $P(X)$ can be achieved through the following steps:
 \begin{itemize}
 \item[$-$] Start from a given warp factor $a(y)$, compute $H(y)$ and use \cref{Feq1,Feq2} to express the pressure as 
 \begin{equation}
 p_5(y)=-\frac3{\kappa_5^2}(H'+2H^2)(y).
 \end{equation}
 For a monotonic warp factor $a(y)$, compute the inverse function $\tilde p_5 : p \mapsto \tilde p_5 (p)$ which satisfies  $\tilde p_5 \circ p_5(y)=y$.
  \item[$-$] Express the density as a function of the pressure by combining the Friedmann equations \labelcref{Feq1,Feq2}:
\begin{align}
\rho_5&=-\frac{3}{\kappa_5^2}H'-p_5\nonumber\\
&=-\frac 3{\kappa_5^2} \left(\frac{a''a-{a'}^2}{a^2}\circ \tilde p_5\right) (p_5) - p_5 \nonumber\\
&\equiv F(-p_5).
\end{align}
\item[$-$] Use the definitions \labelcref{prho1,prho2} to obtain the first order differential equation for $P$
\begin{equation}
2XP_X+F(P)-P=0, \label{reconstruct}
\end{equation}
the solution of which gives the ensemble of functions $P(X)$ giving the warp factor $a(y)$.
 \end{itemize}
 
The above reconstruction method works in principle for any regular enough warp factor. We expect that bulk singularities in the  warp factor would be related to singularities in $P$. Nevertheless, we are naturally interested in regular solutions, which are thus expected to be reconstructed through the inversion of the warp factor profile, as explained above. Among all these regular solutions, the main body of our paper focused on the ones achieving self-tuning of the four-dimensional cosmological constant.

\section{Scalar field junction conditions of non-linear bulk fluids} \label{appendix:B}

This appendix is quite independent and aims at giving a concrete example of the use of k-essence bulk scalars to implement specific bulk fields. As evoked in \cref{appendix:A} they can, in particular, describe the non-linear bulk fluids studied in \cite{Antoniadis:2021grg,Antoniadis:2021rxw}, which lead to regular warp factor profiles. Such fluids satisfy an equation of state given in \cref{nonlinearfluideos}, that we recall here for simplicity
\begin{equation}
p_5=\ell\rho_5^{\lambda}. \label{nonlinearfluideos2}
\end{equation}
The new ingredient with respect to the generic fluid description lies in the scalar field junction conditions discussed in \cref{subsec:scalarjunctionconditions}. In rest of this appendix, we show the explicit form of these junction conditions for generic non-linear bulk fields and for the special case $\lambda=3/2$, solved analytically in \cite{Antoniadis:2021grg,Antoniadis:2021rxw}. We also show explicitly that a self-tuning mechanism cannot be achieved with this type of fluid.

\paragraph{Junction conditions for the general case}
We first study the general case with arbitrary coefficients $\ell,\lambda$ in \cref{nonlinearfluideos2}. It can be described by a k-essence scalar with defining function shown in \eqref{nonlinearPX}, and the expressions appearing in the scalar junction conditions \cref{junction2,junction1} read
\begin{align}
P-2XP_X&=\rho_5=(-\ell)^{\frac 1{1-\lambda}}\left(1+(-X)^{\frac{\lambda-1}{2\lambda}}\right)^{\frac 1{\lambda-1}}\nonumber\\
&\hphantom{=\rho_5\,\,} \equiv B,\label{B}\\
-8XP_X^2&=2(\ell^2)^{\frac 1{1-\lambda}}(-X)^{-\frac{1}{\lambda}}\left(1+(-X)^{\frac{\lambda-1}{2\lambda}}\right)^{\frac 2{\lambda-1}}\nonumber\\
&\equiv A.\label{A}
\end{align}
For a given brane scalar potential $V_{bn}(\phi)$, the scalar field profile is thus determined by solving the equation of motions with boundary conditions corresponding to the junction conditions \labelcref{junction2,junction1} expressed in terms of the above expressions. The scalar field profile will then give back the warp factor profile found in \cite{Antoniadis:2021grg,Antoniadis:2021rxw}. 

\paragraph{Self-tuning mechanism in the general case}
Nevertheless, once the equation of state is fixed, general brane scalar potentials do not lead to self-tuning mechanisms. Generalizing what was studied in \cref{requirementfromjunction}, the self-tuning mechanism is achieved when the brane scalar potential solves the differential equation derived by relating the junction conditions one to the other. We thus relate the two terms of \cref{B,A} through
\begin{equation}
B=(-\ell)^{\frac1{1-\lambda}}\left(1-\frac1{|\ell|}\left(\frac A2\right)^{\frac{1-\lambda}2}\right)^{\frac1{1-\lambda}}\equiv g(A).
\end{equation}
From \cref{junction1,junction2}, we see that the curvature of the brane would thus be expressed as
\begin{align}
\frac{6h^2}{\kappa_5^2a^2}=B-\frac{\kappa_5^2}6V^2_{bn}&=g(A)-\frac{\kappa_5^2}6V^2_{bn}\nonumber\\
&=g(V_{{ bn}, \,\phi}^2)-\frac{\kappa_5^2}6 V^2_{bn},
\end{align}
Hence, the junction conditions impose flat brane solutions, $h=0$, if the brane scalar potential solves the following differential equation
\begin{equation}
g(V_{{bn}, \,\phi}^2)=\frac{\kappa_5^2}{6} V^2_{bn}\,,
\end{equation}
that we rewrite as
\begin{equation}
 {V_{{ bn}, \,\phi}}=\pm \sqrt{2}|\ell|^{\frac1{1-\lambda}}\left|1+\frac 1\ell \left(\frac{\kappa_5^2}6 V^2_{bn}\right)^{{1-\lambda}}\right|^{\frac1{1-\lambda}}. \label{diffequationgeneric}
\end{equation}
This differential equation is obviously separable and, for generic $\lambda$,  a solution can be expressed through the relation
\begin{equation}
\pm \sqrt{2}|\ell|^{\frac1{1-\lambda}} \phi + c_1=\int^{V_{bn}}\! \left|1+\frac 1\ell \left(\frac{\kappa_5^2}{6} V^2\right)^{{1-\lambda}}\right|^{\frac1{\lambda-1}} dV
\end{equation}

This integral can be formally expressed in terms of the hypergeometric function ${}_2F_1$. This leads to the solution
\begin{equation}
\pm \sqrt{2}|\ell|^{\frac1{1-\lambda}} \phi + c_1= V_{bn} \, {}_2F_1\left(2\mathfrak a, \mathfrak a,\mathfrak a+1, z\right), \label{hypergeo}
\end{equation}
where we defined
\begin{equation}
z\equiv-\frac 1\ell \left(\frac{\kappa_5^2}6V_{bn}^2\right)^{{1-\lambda}}, \hspace{0.2cm} \mathfrak{a}=\frac{1}{2(1-\lambda)}. \label{zaparameters}
\end{equation}
The primitive is found using the property
\begin{align}
\forall a,b \quad {\displaystyle {\frac{d}{dz}}\ {}_{2}F_{1}(\mathfrak a,\mathfrak b;\mathfrak a+1;z)={\frac{d}{dz}}\ {}_{2}F_{1}(\mathfrak b,\mathfrak a;\mathfrak a+1;z)}\nonumber\\
{={\frac{\mathfrak a((1-z)^{-\mathfrak b}-{}_{2}F_{1}(\mathfrak a,\mathfrak b;\mathfrak a+1;z))}{z}}}. \label{eqbeforediscussion}
\end{align}
Self-tuning of the brane cosmological constant is achieved when the solution of the differential equation \eqref{diffequationgeneric} is proportional to the integration constant $c_1$, then identified as the brane effective vacuum energy. One should thus invert \cref{hypergeo} to express $V_{bn}$ as a function of $\phi$ to see if self-tuning takes place. Due to the specific form of the \cref{hypergeo}, self-tuning cannot be achieved for non-linear fluids. We explicitly show it in two particular cases below.

\paragraph{Particular values of $\lambda$} For $\lambda=3/2$, the differential equation \eqref{diffequationgeneric} can be simply written as
\begin{equation}
 {V_{{ bn}, \,\phi}}=\pm \frac{\sqrt{2}}{\ell^2}\left(1+\frac{\sqrt{6}}{\ell\kappa_5 |V_{bn}|}\right)^{-2}, \label{diffnonlinearfluidlambda32}
\end{equation}
which is a separable differential equation. After integration, one obtains the following solution 
\begin{align}
\pm \frac{\sqrt{2}}{\ell^2} \phi +c_1=& \frac{2\sqrt{6}}{\ell\kappa_5} \log(|V_{bn}|)-\frac6{\ell^2\kappa_5^2 |V_{bn}|} + |V_{bn}|.
\end{align}
This relation should be inverted to express $V_{bn}$ as a function of $\phi$, but it is clear that this function depends non-linearly on $c_1$.

For $\lambda=2$, the $z$ and $\mathfrak a$ parameters, introduced in \cref{zaparameters}, and the right-hand side of \cref{hypergeo} read
\begin{align} 
&\mathfrak a=-1/2, \qquad z=-\frac 1\ell \left(\frac{\kappa_5^2}6V_{bn}^2\right)^{-1}, \\
&V_{bn} \, {}_2F_1\left(2\mathfrak a,\mathfrak a,\mathfrak a+1, z\right)=\frac{-6 + V_{bn}^2 \ell \kappa_5^2}{V_{bn} \ell \kappa_5^2},
\end{align}  
so that \cref{hypergeo} is a binomial solved by
\begin{align}
\hspace{0.5cm}V_{bn}(\phi)=&\frac{c_1}{2}\pm\frac{|\ell|^{\frac{1}{1-\lambda}}}{\sqrt{2}}\phi  \label{case1}\\
&\pm \frac{1}{\sqrt{|\ell|} \kappa_5}{\sqrt{ 24 + \ell\kappa_5^2 \left(c_1 \pm 
      \sqrt{2} |\ell|^{\frac1{1 - \lambda}} \phi\right)^2}}.\nonumber
\end{align}
 In this case, we see explicitly that $c_1$ does not factorize. 
 
 Hence, according to the discussion after \cref{eqbeforediscussion}, self-tuning of the cosmological constant cannot be achieved in either of the above cases.

\bibliography{CuscutonBraneW}
\bibliographystyle{JHEP}
   
\end{document}